\documentclass[12pt]{article}
\usepackage{amsfonts,latexsym,graphicx,amsmath,cite}

\usepackage{comment} 
\usepackage[usenames,dvipsnames]{color}
\usepackage{amssymb,amsmath,amscd}
\usepackage{dsfont}
\usepackage{amsthm}
\usepackage{multirow}
\usepackage{verbatim}   
\usepackage{amsfonts}     
 \usepackage{pifont} 
\usepackage[color,all,cmtip]{xy}

\def\a{\alpha}
\def\b{\beta}

\def\g{\gamma}
\def\k{\kappa}

\def\s{\sigma}

\newcommand{\eea}[1]{\begin{align}#1\end{align}}
\newcommand{\eean}[1]{\begin{align*}#1\end{align*}}
\def\ba{\begin{array}}
\def\ea{\end{array}}

\def\dag{\dagger}
\def\nn{\nonumber}

\def\dag{^\dagger}

\newcommand{\reff}[1]{(\ref{#1})}

\newcommand{\ket}[1]{|#1\rangle}

\newcommand{\conj}[1]{\overline{#1}}

\makeatletter
\renewcommand{\paragraph}{\@startsection{paragraph}{4}{0ex}%
   {-3.25ex plus -1ex minus -0.2ex}%
   {1.5ex plus 0.2ex}%
   {\normalfont\normalsize\bfseries}}
\makeatother

\theoremstyle{plain}

\begin{document}

\title{A variational approach for the Quantum Inverse Scattering Method}
\date{}
\author{A. Birrell, P.S. Isaac and J. Links\\
Centre for Mathematical Physics \\
School of Mathematics and Physics \\
The University of Queensland, 4072 \\
Australia}
  
  \maketitle
\begin{abstract} 
We introduce a variational approach for the Quantum Inverse Scattering Method to exactly
solve a class of Hamiltonians via Bethe ansatz methods. We undertake this in a manner
which does not rely on any prior knowledge of integrability through the existence of a set
of conserved operators. The procedure is conducted in the framework of Hamiltonians
describing the crossover between the low-temperature phenomena of superconductivity,
in the Bardeen-Cooper-Schrieffer (BCS) theory, and Bose-Einstein
condensation (BEC). The Hamiltonians considered describe systems with 
interacting Cooper pairs and a bosonic degree
of freedom. We obtain general exact solvability requirements which include
seven subcases which have previously appeared in the literature.
\end{abstract}

\section{Introduction}

The Quantum Inverse Scattering Method (QISM) \cite{stf79,tf79,ks79} is used for
constructing quantum Hamiltonians with multiple conserved operators, and in turn obtaining
their exact solutions by Bethe ansatz methods. A solution of the Yang-Baxter equation
\cite{m64,y67,b72} may be used to construct a {\it transfer matrix} which generates the
conserved operators of the Hamiltonian. The Bethe ansatz used to obtain the exact solution
can assume many forms. The original approach due to Bethe \cite{bethe} is commonly termed
the co-ordinate Bethe ansatz, whereas a more modern approach in the framework of the QISM
(subject to a suitable reference state) is the algebraic Bethe ansatz
\cite{stf79,tf79,ks79}. However there are several variants between these two formats,
which rely on functional relations and assumed analyticity properties of the eigenvalue
spectrum to obtain the exact solution  \cite{r83,bbp90,wz94,n03,cdil10}. We note that in
principle the implementation of the co-ordinate Bethe ansatz is not dependent on any prior
knowledge of an associated solution of the Yang-Baxter equation, nor the conserved
operators of the Hamiltonian that it generates.

Progress in cold atom physics has yielded many studies into the nature of the crossover 
between the low-temperature phenomena of superconductivity
from the Bardeen-Cooper-Schrieffer (BCS) theory and Bose-Einstein
condensation (BEC) \cite{rgj04,gr07,bdz08,lcch10}. Early theoretical accounts emphasised the need
to study Hamiltonians which explicitly incorporate coupling between Cooper pairs of atoms
and bosonic molecular modes \cite{hkcw01,og02}. The goal of the work presented here is to
implement a strategy motivated by both  the co-ordinate and algebraic Bethe ansatz
approaches to obtain a general class of exactly solvable Hamiltonians, with both a bosonic
and Cooper pairing degrees of freedom, such that they model BCS-BEC crossover behaviour.
As we will show, this approach gives a unified construction for classes of exactly solvable Hamiltonians
with multiple free coupling parameters. It reproduces some seven exactly solvable subcases  
which have previously appeared in the literature (including some for which the bosonic degree of freedom 
is decoupled from the system, leaving a BCS Hamiltonian) 
\cite{r63,g76,dhl06,ilsz09,dilz11,lrdo11,dlrrr11}.  The availability of such exact results
makes amenable the computation of correlation functions \cite{dilz11,lrdo11,dilsz10} which
potentially can be compared to experimental results.

To highlight the {\it mathematical} motivation for our work, it is instructive to examine
the evolution of integrable models of correlated electrons in one-dimension. Here there
are many examples whereby models were first introduced and solved exactly via the
co-ordinate Bethe ansatz approach, with the rederivation of the Hamiltonian through the
Yang-Baxter equation coming as a later development. The first of these is the Hubbard
model which was exactly solved by Lieb and Wu in 1968 \cite{lw68}. It was not until 1986
that Shastry \cite {s86} first made connection to the Yang-Baxter equation for this model,
with subsequent understanding of the algebraic structure developing through the 1990s
(e.g. \cite{rm97,usw98}). The strong coupling limit of the Hubbard model leads to the
$t$-$J$ model, which was shown to be exactly solvable for particular choices of the coupling
parameters through the works of Schlottmann \cite{s87} and Sarkar \cite{s90}. In this case
it was not so long before the conserved operators were constructed through the Yang-Baxter
equation \cite{ek92,fk93}. Following on from this there was a flourish of activity
throughout the 1990s in the study of exactly solvable models for correlated electrons via
co-ordinate Bethe ansatz, which included the Bariev model \cite{b91}, the $q$-deformation
of the supersymmetric $U$ model \cite{bkz95}, and the Alcaraz-Bariev model which
ultimately included both Hubbard and a solvable $t$-$J$ model as particular subcases
\cite{ab99}. All three Hamiltonians were later shown to be derivable from a solution of
the Yang-Baxter equation. Some works showing the connection to the Yang-Baxter equation,
viz.  \cite{z96,sw97} for the Bariev model, and \cite{ghlz96} for the $q$-deformed
supersymmetric $U$ model, were reasonably rapid developments. The algebraic formulations
for the Bethe ansatz solution were given in \cite{mr97,z97,lf01}. Recently the origin of
the solution of the Yang-Baxter equation for the Alcaraz-Bariev model has started to
become clear \cite{bk08}.    

There are two key differences between the construction of correlated electrons referred to above and our approach below. The first is that in the co-ordinate Bethe ansatz approach, the ansatz wave-functions for the correlated electron models are taken to be superpositions of plane waves. We will instead formulate the wave-functions as a factorisable operator acting on a reference state, which follows the spirit of the algebraic Bethe ansatz \cite{stf79,tf79,ks79} and relates back to Richardson's original calculations for pairing Hamiltonians \cite{r63}. However we differ from the algebraic Bethe ansatz approach through a second key point in that we will work directly with a variational Hamiltonian. We follow the spirit of the co-ordinate Bethe ansatz approach \cite{bethe,lw68,s87,s90,b91,bkz95,ab99} in that we do not resort to any knowledge of a transfer matrix or a set of conserved operators.   By combining these two aspects of the co-ordinate and algebraic Bethe ansatz methods, we are able to formulate a powerful technique for constructing exactly solvable models in a very general fashion.         

The structure of the manuscript is as follows.  We begin Section 2 by introducing a general Hamiltonian describing a reduced BCS model coupled to a bosonic degree of freedom. In Section 2.1 we apply the Bethe ansatz to obtain constraints on the Hamiltonian's coupling parameters which are sufficient for exact solvability of the Hamiltonian. There are two cases which are dealt with separately in Section 2.1.1 and Section 2.1.2. In section 3 we discuss the connection between seven known exactly solvable subcases that are limits of the general model we consider.    
   
\section{Variational Hamiltonian}
We consider the general family of pairing Hamiltonians coupled to a bosonic degree of freedom
\eea{\label{genHAM} H = H_0-H_1}
where
\eea{
\label{H0eq} H_0 &= \a N_0 + \k N_0^2+ \sum_{k=1}^L f(z_k)N_k,\\
\label{Teq} H_1 &=  \b \sum_{k=1}^L g(z_k)b_0 b_k\dag + \b\sum_{k=1}^L \conj{g(z_k)} b_0\dag b_k +\s\sum_{k,s}^L g(z_k)\conj{g(z_s)}b_k\dag b_s,
}
for some complex-valued functions $f(z)$ and $g(z),$ and real-valued $\a,$ $\k,$ $\b$ and
$\s,$ which will be subject to certain solvability constraints yet to be determined. The overline notation denotes complex conjugation, which imposes that the above Hamiltonian is hermitian. The operators $b_k\dag=c_{-k}\dag c_{k}\dag$ and $N_k=b_k\dag b_k$ for $k>0$ are hard-core Cooper pair creation and number operators, where $c_k\dag$ are fermion creation operators and Cooper pairs are assumed to consist of paired fermions of zero total momentum. Spin labels have been suppressed, so we do not imply anything about the spin properties (e.g., singlet, triplet) of the pairing. We also have a single bosonic mode with operators $b_0\dag$ and $N_0$. The particle operators satisfy the following commutation relations:
\eea{\label{hccomm}
\begin{split}[b_j,b_k]=& 0 =[b_j\dag,b_k\dag] \qquad\qquad\quad ~~\forall~ j,k\geq 0,\\
[b_j,b_k\dag] =& 0 \qquad\qquad\qquad~~\forall~ j,k\geq 0, ~j\neq k,\\
[b_0,b_0\dag] =& I,~~ [b_k,b_k\dag] = I-2N_k \quad~~\forall~ k> 0.
\end{split}}
The Hamiltonian commutes with the total number operator $N=N_0+\sum_{k} N_k$. Hamiltonians in this family describe a system of bosonic molecules, condensed into a single bosonic degree of freedom, coupled to $L$ Cooper pairs which are bosonic-like Cooper pairs that must observe an exclusion principle. The Hamiltonians consist of two parts. The diagonal part $H_0$, given in equation \reff{H0eq}, describes the bosonic mode and allowed Cooper pair energy levels. The self-interaction term $\k N_0^2$ describes a shift in the frequency of the bosonic mode as it is populated. The cross-interaction part $H_1$, given in equation \reff{Teq}, describes level dependent molecule-pair coupling and pair-pair couplings when either of $\beta $ or $\sigma$ are non-zero. For the special case $\a=\b=\k=0$, which suppresses any action of the Hamiltonian on the bosonic part of the underlying Hilbert space, the Hamiltonian \reff{genHAM} reduces to a general form of the BCS Hamiltonian. It is for this reason we can refer to the general pairing Hamiltonian as a BEC-BCS crossover Hamiltonian.

We do not expect that the Hamiltonian $\reff{genHAM}$ is exactly solvable in general. However, we have found solvability conditions for various sub-classes of the Hamiltonian, in particular the cases $i)~\k=0$ of no self-interaction term and $ii)~\s=0$ of no BCS pair-pair scattering term. Taking appropriate limits, these two cases are shown to reduce consistently to the same Hamiltonian when $\k=0=\s$.

\subsection{Exact Solvability Constraints\label{sec:constraints}}

We would like to understand the extent to which an exact solution can be found for pairing Hamiltonians of the form given in equation \reff{genHAM} for the yet to be determined functions $f(z)$ and $g(z)$. 
To begin, motivated by the approach of  Richardson \cite{r63}, we assume the ansatz\footnote{For simplicity, throughout the manuscript we only consider eigenstates which do not involve {\it blocked} levels. For a review of the blocking effect we refer to \cite{vr01}. There is no technical impediment to extend results to accommodate the general case with blocked levels, but we omit instances with blocked states for the sake of readability.}
\eea{
\label{ansatz} \ket{\Psi} = \prod_{j=1}^M C(y_j)\ket{0}
}
for the eigenstates of \reff{genHAM}, where $\ket{0}$ denotes the vacuum state, 
\eean{
C(y) = \g(y)b_0\dag + \sum_{k=1}^L h(y,z_k)b_k\dag,~~y\in\mathbb{C}
}
and $h(y,z)$ is yet to be determined. We introduce the notation
\eean{
\ket{\Psi_j} = \prod_{l\neq j}^M C(y_l)\ket{0},\hspace{.5in}\ket{\Psi_{ij}} = \prod_{l\neq i,j}^M C(y_l)\ket{0}.
}
Using the identities in \reff{hccomm} the following commutation relations are found:
\eean{\begin{split}
 [b_0,C(y)] &= \g(y)I\\
 \left[b_k,C(y)\right] &= h(y,z_k)(I-2N_k),\\
 \left[N_0,C(y)\right] &= \g(y)b_0\dag,\\
 \left[N_k,C(y)\right] &= h(y,z_k)b_k\dag,\\
 \left[N_0^2,C(y)\right] &= \g(y)b_0\dag(I + 2N_0),\\
 \left[H_0,C(y)\right] &= \a\g(y) b_0\dag + \sum_{k=1}^L f(z_k)h(y,z_k)b_k\dag +\k\g(y)b_0\dag(I+2N_0).
\end{split}}
We note that for an operator $\hat{O}$,
\eean{
\left[\hat{O}, \prod_{j=m}^M C(y_j)\right] = \sum_{l=m}^M\left(\prod_{r=m}^{l-1}C(y_r)\right)[\hat{O},C(y_l)]\prod_{j=l+1}^M C(y_j).
}
Using this identity we can show
\eean{
b_0\ket{\Psi} &= \left[b_0,\prod_{j=1}^MC(y_j)\right]\ket{0} \\
&= \sum_{j=1}^M\g(y_j)\ket{\Psi_j},\\
b_k\ket{\Psi} &= \sum_{j=1}^M\left(\prod_{r=1}^{j-1}C(y_r)\right)\left[b_k,C(y_j)\right]\prod_{l=j+1}^{M}C(y_l)\ket{0}\\
&= \sum_{j=1}^M h(y_j,z_k)\ket{\Psi_j}-\sum_{j,l\neq j}^M h(y_j,z_k)h(y_l,z_k)b_k\dag\ket{\Psi_{jl}}.
}
We may then calculate the following
\eean{
H_0\ket{\Psi} &= (\a+\k)\sum_{j=1}^M \g(y_j)b_0\dag\ket{\Psi_j} + \sum_{j=1}^M \sum_{k=1}^L f(z_k)h(y_j,z_k)b_k\dag\ket{\Psi_j}  +\k\sum_{j,l\neq j}^M \g(y_j)\g(y_l)b_0\dag b_0\dag\ket{\Psi_{jl}},\\
\begin{split}H_1\ket{\Psi} &= \b\sum_{j=1}^M \sum_{k=1}^L \g(y_j)g(z_k) b_k\dag\ket{\Psi_j} +\b\sum_{j=1}^M \sum_{k=1}^L\conj{g(z_k)}h(y_j,z_k)b_0\dag\ket{\Psi_j}\\
& -\b\sum_{j,l\neq j}^M \sum_{k=1}^L \conj{g(z_k)}h(y_j,z_k)h(y_l,z_k) b_0\dag b_k\dag\ket{\Psi_{jl}} +\s\sum_{j=1}^M \sum_{k,s=1}^L g(z_k)\conj{g(z_s)}h(y_j,z_s)b_k\dag\ket{\Psi_j}\\
& -\s\sum_{j,l\neq j}^M \sum_{k,s=1}^Lg(z_k)\conj{g(z_s)}h(y_j,z_s)h(y_l,z_s) b_k\dag b_s\dag\ket{\Psi_{jl}}.\end{split}
}
These equations give the action of the Hamiltonian $H$ on the state $\ket{\Psi}$. In order to determine the exact solution we require that $\ket{\Psi}$ be an eigenstate of $H$. Thus we look to solve the eigenvalue problem:
\eea{
\label{GenEval} (H_0-H_1)\ket{\Psi} &= E\ket{\Psi}
}
for some scalar $E$. In general this will not be possible. The set of constraints required to find a solution to equation \reff{GenEval} are called solvability constraints and define the manifold in the coupling parameter space along which the Hamiltonian \reff{genHAM} sustains an exact solution.

In order to find an exact solution we look to write the sums involving any of the vectors $b_k\dag b_s\dag\ket{\Psi_{jl}}$, $b_0\dag b_k\dag\ket{\Psi_{jl}}$, $b_0\dag b_0\dag\ket{\Psi_{jl}}$, $b_k\dag\ket{\Psi_{j}}$, or $b_0\dag\ket{\Psi_{j}}$ as a combination of the vectors $\ket{\Psi_{j}}$ and $\ket{\Psi}$. As a first step we write the sum over $b_k\dag b_s\dag\ket{\Psi_{jl}}$ terms as a sum over a $b_0\dag b_k\dag\ket{\Psi_{jl}}$ part and a $b_k\dag\ket{\Psi_j}$ part. However, comparing the $b_k\dag b_s\dag\ket{\Psi_{jl}}$ and $b_0\dag b_k\dag\ket{\Psi_{jl}}$ terms we see that the particular constraint chosen to reduce the $b_k\dag b_s\dag\ket{\Psi_{jl}}$ terms must be compatible with one chosen to reduce the $b_0\dag b_k\dag\ket{\Psi_{jl}}$ terms. For compatibility we introduce the constraint
\eean{
\conj{g(z_s)}h(y_j,z_s)h(y_l,z_s) = k(y_j,y_l)h(y_l,z_s) +k(y_l,y_j)h(y_j,z_s)
,~~\forall y_l,y_j,z_s\in \mathbb{C}
}
where $k(y_l,y_j)$ is to be chosen later. This allows us to write
\eean{
\begin{split}H_1\ket{\Psi} &= \b\sum_{j=1}^M \sum_{k=1}^L \g(y_j)g(z_k) b_k\dag\ket{\Psi_j} +\b\sum_{j=1}^M \sum_{k=1}^L\conj{g(z_k)}h(y_j,z_k)b_0\dag\ket{\Psi_j}\\
& -2\b\sum_{j,l\neq j}^M \sum_{k=1}^L k(y_l,y_j)h(y_j,z_k)b_0\dag b_k\dag\ket{\Psi_{jl}} +\s\sum_{j=1}^M \sum_{k,s}^L g(z_k)\conj{g(z_s)}h(y_j,z_s)b_k\dag\ket{\Psi_j}\\
& -2\s\sum_{j,l\neq j}^M \sum_{k,s}^Lg(z_k)k(y_l,y_j)h(y_j,z_s) b_k\dag b_s\dag\ket{\Psi_{jl}}.\end{split}
}
We then utilise the definition \reff{ansatz} for $\ket{\Psi}$ to express the sums in terms of the desired vectors,
\eean{
\begin{split}H_1\ket{\Psi} &= \b\sum_{j=1}^M \sum_{k=1}^L \g(y_j)g(z_k) b_k\dag\ket{\Psi_j} +\b\sum_{j=1}^M \sum_{k=1}^L\conj{g(z_k)}h(y_j,z_k)b_0\dag\ket{\Psi_j} -2\b\sum_{j,l\neq j}^M k(y_j,y_l)b_0\dag \ket{\Psi_{j}}\\
& +2\b\sum_{j,l\neq j}^M k(y_j,y_l)\g(y_l)b_0\dag b_0\dag\ket{\Psi_{jl}} +\s\sum_{j=1}^M \sum_{k,s}^L g(z_k)\conj{g(z_s)}h(y_j,z_s)b_k\dag\ket{\Psi_j}\\
& -2\s\sum_{j,l\neq j}^M \sum_{k=1}^Lg(z_k)k(y_j,y_l)b_k\dag\ket{\Psi_{j}}+2\s\sum_{j,l\neq j}^M \sum_{k=1}^Lg(z_k)k(y_j,y_l)\g(y_l) b_0\dag b_k\dag\ket{\Psi_{jl}}.\end{split}
}
Consider the full expression
\eean{
H\ket{\Psi} &= (\a+\k)\sum_{j=1}^M \g(y_j)b_0\dag\ket{\Psi_j} -\b\sum_{j=1}^M \sum_{k=1}^L\conj{g(z_k)}h(y_j,z_k)b_0\dag\ket{\Psi_j}  +2\b\sum_{j,l\neq j}^M k(y_j,y_l)b_0\dag \ket{\Psi_{j}}\\
& +\sum_{j=1}^M \sum_{k=1}^L f(z_k)h(y_j,z_k)b_k\dag\ket{\Psi_j} -\b\sum_{j=1}^M \sum_{k=1}^L \g(y_j)g(z_k) b_k\dag\ket{\Psi_j}   -\s\sum_{j=1}^M \sum_{k,s}^L g(z_k)\conj{g(z_s)}h(y_j,z_s)b_k\dag\ket{\Psi_j}\\
& +2\s\sum_{j,l\neq j}^M \sum_{k=1}^Lg(z_k)k(y_j,y_l)b_k\dag\ket{\Psi_{j}}-2\s\sum_{j,l\neq j}^M \sum_{k=1}^Lg(z_k)k(y_j,y_l)\g(y_l) b_0\dag b_k\dag\ket{\Psi_{jl}}\\
& -2\b\sum_{j,l\neq j}^M k(y_j,y_l)\g(y_l)b_0\dag b_0\dag\ket{\Psi_{jl}}  +\k\sum_{j=1,l\neq j}^M \g(y_j)\g(y_l)b_0\dag b_0\dag\ket{\Psi_{jl}}.
}
At this point we observe the appearance of identical coefficients of the vectors $\b b_0\dag\ket{\Psi_j}$ and $\s\sum_{k=1}^Lg(z_k)b_k\dag\ket{\Psi_j}$ for $\s\neq0\neq\b$. In finding the exact solution we must choose constraints on the coefficients that are compatible with this. Thus, the next constraint must involve the terms that are not yet related. We seek to express the remaining sums over vectors $b_k\dag\ket{\Psi_j}$ as a combination of the vectors $\ket{\Psi}$, $\b b_0\dag\ket{\Psi_j}$ and $\s\sum_{k=1}^Lg(z_k)b_k\dag\ket{\Psi_j}$ such that the latter two have the same coefficients. We choose the following constraint to reduce the terms to the desired vectors 
\eea{
f(z_k)h(y_j,z_k)&= y_jh(y_j,z_k)+g(z_k)r(y_j) \label{ConstraintChoice} \\
\Rightarrow h(y_j,z_k) &= \frac{g(z_k)r(y_j)}{f(z_k)-y_j} ~~\forall y_j,z_k\in\mathbb{C}, \nonumber
}
where the $r(y_j)$ will be determined by compatibility requirements. With these choices we find
\eean{
\sum_{j=1}^M \sum_{k=1}^L f(z_k)h(y_j,z_k)b_k\dag\ket{\Psi_j} &= 
\sum_{j=1}^M \sum_{k=1}^L y_jh(y_j,z_k)b_k\dag\ket{\Psi_j} +\sum_{j=1}^M \sum_{k=1}^L g(z_k)r(y_j)b_k\dag\ket{\Psi_j}\\
&= 
\sum_{j=1}^M y_j\ket{\Psi} -\sum_{j=1}^M y_j\g(y_j)b_0\dag\ket{\Psi_j} +\sum_{j=1}^M \sum_{k=1}^L g(z_k)r(y_j)b_k\dag\ket{\Psi_j}.
}
Use of this relation leads to 
\eean{
H\ket{\Psi} &= \sum_{j=1}^M y_j\ket{\Psi} +\sum_{j=1}^M (\a+\k-y_j)\g(y_j)b_0\dag\ket{\Psi_j} -\b\sum_{j=1}^M \sum_{k=1}^L\conj{g(z_k)}h(y_j,z_k)b_0\dag\ket{\Psi_j}\\
&  +2\b\sum_{j,l\neq j}^M k(y_j,y_l)b_0\dag \ket{\Psi_{j}} +\sum_{j=1}^M \sum_{k=1}^L g(z_k)[r(y_j)-\b\g(y_j)]b_k\dag\ket{\Psi_j}\\
&   -\s\sum_{j=1}^M \sum_{k,s}^L g(z_k)\conj{g(z_s)}h(y_j,z_s)b_k\dag\ket{\Psi_j} +2\s\sum_{j,l\neq j}^M \sum_{k=1}^Lg(z_k)k(y_j,y_l)b_k\dag\ket{\Psi_{j}}\\
&  -2\b\sum_{j,l\neq j}^M k(y_j,y_l)\g(y_l)b_0\dag b_0\dag\ket{\Psi_{jl}}  +\k\sum_{j=1,l\neq j}^M \g(y_j)\g(y_l)b_0\dag b_0\dag\ket{\Psi_{jl}}\\
& -2\s\sum_{j,l\neq j}^M \sum_{k=1}^Lg(z_k)k(y_j,y_l)\g(y_l) b_0\dag b_k\dag\ket{\Psi_{jl}}.
}
At this stage we remark that the choice of constraint \reff{ConstraintChoice} sets the
energy eigenvalues to a standardised form $\displaystyle{E = \sum_{j=1}^M y_j.}$ 

Exact solvability is achieved by requiring that the coefficients of all terms other than $\ket{\Psi}$ cancel. To meet this requirement we choose (for $g(z_k)\neq0$)
\eea{
\nn (y_j-\a-\k)\g(y_j)  +\b\sum_{k=1}^L\conj{g(z_k)}h(y_j,z_k) &= 2\b\sum_{l\neq j}^M k(y_j,y_l)\\
\nn \b\g(y_j) -r(y_j) +\s \sum_{k=1}^L \conj{g(z_k)}h(y_j,z_k) & =2\s\sum_{l\neq j}^Mk(y_j,y_l)\\
\label{const4} \b\left(k(y_j,y_l)\g(y_l)+k(y_l,y_j)\g(y_j)\right) &= \k\g(y_j)\g(y_l)\\
\label{const5} \s\left(k(y_j,y_l)\g(y_l) +k(y_l,y_j)\g(y_j)\right) &= 0.
}
However, we note for $\k\neq 0 \neq\s$ the constraints \reff{const4} and \reff{const5} are
incompatible and we have at least two separate cases. We must now derive two sets of
solvability conditions, one for each case, however, they should agree in the appropriate
limit $\k=0=\s$ when they describe the same family of Hamiltonians. Some care will also be
needed in the limit $\b\to0$.

\subsubsection{Case 1: No Pair-Pair Interaction ($\s=0$)\label{sec:case1}}
The first case we look at is $\s=0$. Here the solvability conditions reduce to  
\eea{
\label{c1} \conj{g(z_k)}h(y_j,z_k)h(y_l,z_k) &= k(y_j,y_l)h(y_l,z_k) +k(y_l,y_j)h(y_j,z_k)\\
\label{c2} (y_j-\a-\k)\g(y_j) +\b\sum_{k=1}^L\conj{g(z_k)}h(y_j,z_k) &= 2\b\sum_{l\neq j}^M k(y_j,y_l)\\
\label{c3} \b\g(y_j)-r(y_j) & = 0\\
\label{c4}\b\left(k(y_j,y_l)\g(y_l)+k(y_l,y_j)\g(y_j)\right) &= \k\g(y_j)\g(y_l)\\
\label{c5} h(y_j,z_k) &= \frac{g(z_k)r(y_j)}{f(z_k)-y_j}.
}
Firstly, \reff{c3} is satisfied for non-trivial $\g(y_j)$ only if
\eean{
\b\g(y_j) = r(y_j).
}
Substituting \reff{c3} into \reff{c5} gives
\eean{
h(y_j,z_k) &= \frac{\b g(z_k)\g(y_j)}{f(z_k)-y_j}
}
and substituting this into \reff{c1} gives
\eean{
\conj{g(z_k)}\frac{\b g(z_k)\g(y_j)}{f(z_k)-y_j}\frac{\b g(z_k)\g(y_l)}{f(z_k)-y_l} &= k(y_j,y_l)\frac{\b g(z_k)\g(y_l)}{f(z_k)-y_l} +k(y_l,y_j)\frac{\b g(z_k)\g(y_j)}{f(z_k)-y_j}
}
Rearranging,
\eean{
f(z_k)\frac{\b(k(y_j,y_l)\g(y_l)+k(y_l,y_j)\g(y_j))}{\g(y_j)\g(y_l)} - \conj{g(z_k)}g(z_k) &= \frac{\b(k(y_j,y_l)\g(y_l)y_j +k(y_l,y_j)\g(y_j)y_l)}{\g(y_j)\g(y_l)}
}
Since the right hand side does not depend on the parameter $z_k$ we require that, for some constants $c_1$ and $c_2$,
\eean{
\b(k(y_j,y_l)\g(y_l)y_j +k(y_l,y_j)\g(y_j)y_l) &= c_1 \g(y_j)\g(y_l),\\
\b(k(y_j,y_l)\g(y_l) +k(y_l,y_j)\g(y_j)) &= c_2 \g(y_j)\g(y_l),\\
c_2 f(z_k)-\b^2g(z_k)\conj{g(z_k)} &= c_1.
}
However, compatibility with constraint \reff{c4} requires
\eean{
c_2 = \k.
}
The solution is thus
\eean{
k(y_j,y_l) &= \frac{c_1-c_2 y_l}{\b(y_j-y_l)} \g(y_j),\\
f(z_k) &= c_2^{-1} \left(\b^2 g(z_k)\conj{g(z_k)}+c_1\right),\\
c_2 &=\k,
}
for some constant $c_1$.

Substituting this into the constraint \reff{c2} completes the compatibility of constraints yielding 
\eea{
\label{intmans0a} y_j-(\a+\k) +\sum_{k=1}^L\frac{c_2 f(z_k)- c_1}{f(z_k)-y_j} &= 2\sum_{l\neq j}^M \frac{c_1-c_2 y_l}{y_j-y_l}\\
\nn c_2&=\k.
}
We refer to \reff{intmans0a} as the Bethe ansatz equations for which the sub-family of Hamiltonians of \reff{genHAM} corresponding to $\s=0$ are solvable. An equivalent expression is obtained by manipulating the terms,
\eea{\label{intmans0b}
\begin{split}
\frac{1}{2y_j}(\a+\k+c_2(2M-2-L)) -\frac{1}{2}  +\frac{1}{4}\sum_{k=1}^L\frac{2}{y_j/c_2-f(z_k)/c_2}&\\ +\frac{c_1}{2y_j}\left(\sum_{k=1}^L\frac{1}{f(z_k)-y_j}  +\sum_{l\neq j}^M \frac{2}{y_j-y_l}\right)&=  \sum_{l\neq j}^M \frac{c_2}{y_j-y_l}
\end{split}\\
\nn c_2&=\k.
}
In the special case of $\k=0=\s$ we find $c_1 = -\b^2|g(z_k)|^2=-G^2,$ $c_2=0$ and the Bethe ansatz equations reduce to
\eean{
\frac{\a}{2G} -\frac{y_j/G}{2} -\frac{1}{4}\sum_{k=1}^L\frac{2}{f(z_k)/G-y_j/G} &= \frac{1}{2}\sum_{l\neq j}^M \frac{2}{y_j/G-y_l/G}.
}

We summarise by listing the constraining relations for exact solvability in the case $\s=0$ here:
\eean{
 h(y_j,z_k) &= \frac{\b g(z_k)\g(y_j)}{f(z_k)-y_j}\\
 f(z_k) &= \k^{-1}\b^2 g(z_k)\conj{g(z_k)} +\k^{-1}c_1\\
 y_j-(\a+\k) +\sum_{k=1}^L\frac{c_2 f(z_k)- c_1}{f(z_k)-y_j} &= 2\sum_{l\neq j}^M \frac{c_1-c_2 y_l}{y_j-y_l}\\
c_2&=\k.
}
for constants $\b$, and $c_1$. These constraints define the manifold in the coupling parameters of the pairing Hamiltonian \reff{genHAM} for which it is exactly solvable with eigenstates of the form \reff{ansatz} in the case $\s=0$. In the above derivation we find that $\g(y_j)$ is not fixed, however, it will be fixed by normalisation of the eigenstate.

\subsubsection{Case 2: No Self-Interaction Term ($\k=0$)\label{sec:case2}}
Setting $\k=0$ in the solvability conditions leads to the following set of constraints,
\eea{
\label{c21} \conj{g(z_k)}h(y_j,z_k)h(y_l,z_k) &= k(y_j,y_l)h(y_l,z_k) +k(y_l,y_j)h(y_j,z_k)\\
\label{c22} (y_j-\a)\g(y_j) +\b\sum_{k=1}^L\conj{g(z_k)}h(y_j,z_k) &= 2\b\sum_{l\neq j}^M k(y_j,y_l)\\
\label{c23}  \b\g(y_j)-r(y_j) +\s \sum_{k=1}^L \conj{g(z_k)}h(y_j,z_k) & =2\s\sum_{l\neq j}^Mk(y_j,y_l)\\
\label{c24} k(y_j,y_l)\g(y_l)+k(y_l,y_j)\g(y_j) &= 0\\
\label{c25} h(y_j,z_k) &= \frac{g(z_k)r(y_j)}{f(z_k)-y_j}.
}

Constraints \reff{c22} and \reff{c23} are compatible when 
\eean{
\b r(y_j) = [\s(\a-y_j)+\b^2]\g(y_j)
}
which along with condition \reff{c24} results in the constraint
\eea{
\label{c24a} 
\b[\s(\a-y_j)+\b^2]k(y_j,y_l)r(y_l)+\b[\s(\a-y_l)+\b^2]k(y_l,y_j)r(y_j) &= 0.
}
and along with \reff{c25} results in 
\eean{
 \b h(y_j,z_k) &= [\s(\a-y_j)+\b^2]\frac{g(z_k)\g(y_j)}{f(z_k)-y_j}
}
Substituting \reff{c25} into \reff{c21} gives
\eean{
\conj{g(z_k)}\frac{g(z_k)r(y_j)}{f(z_k)-y_j}\frac{g(z_k)r(y_l)}{f(z_k)-y_l} &= k(y_j,y_l)\frac{g(z_k)r(y_l)}{f(z_k)-y_l} +k(y_l,y_j)\frac{g(z_k)r(y_j)}{f(z_k)-y_j}
}
which we rearrange to
\eean{
f(z_k)\frac{k(y_j,y_l)r(y_l)+k(y_l,y_j)r(y_j)}{r(y_j)r(y_l)} -\conj{g(z_k)}g(z_k) &= \frac{k(y_j,y_l)r(y_l)y_j +k(y_l,y_j)r(y_j)y_l}{r(y_j)r(y_l)}.
}
Since the right hand side of the equation does not depend on the parameter $z_k$ we must satisfy, for some constants $c_2$ and $c_1$, the following set of relations:
\eean{
 \begin{split} k(y_j,y_l)r(y_l)y_j +k(y_l,y_j)r(y_j)y_l &= c_1 r(y_j)r(y_l),\\
k(y_j,y_l)r(y_l)+k(y_l,y_j)r(y_j) &= c_2 r(y_j)r(y_l),\\
c_2 f(z_k)-\conj{g(z_k)}g(z_k)& = c_1.\end{split}
}
For compatibility of the first two equations we find the solution must be of the form
\eean{
k(y_j,y_l)&= \frac{c_1 -c_2y_l}{y_j -y_l}r(y_j).
}
However, for compatibility with constraint \reff{c24a} we then require
\eean{
\b[\s(\a-y_j)+\b^2]\frac{c_1 -c_2y_l}{y_j -y_l}r(y_j)r(y_l)-\b[\s(\a-y_l)+\b^2]\frac{c_1 -c_2y_j}{y_j -y_l}r(y_l)r(y_j) &= 0
}
which we can rearrange to obtain
\eean{
\b[c_2(\s\a+\b^2)-c_1\s]r(y_j)r(y_l) &= 0
}
and we must have
\eea{
\label{param2constraint}  c_2(\s\a+\b^2)\b =  c_1 \s\b.
}
At this point we have the conditions
\eean{
c_1 \s\b &= c_2(\s\a+\b^2)\b ,\\
k(y_j,y_l) &= \frac{c_1 -c_2y_l}{y_j -y_l}r(y_j),\\
c_1 &= c_2 f(z_k)-\conj{g(z_k)}g(z_k),\\
\b r(y_j) &= [\s(\a-y_j)+\b^2]\g(y_j),\\
h(y_j,z_k) &= \frac{g(z_k)r(y_j)}{f(z_k)-y_j}.
}
For $\b\neq 0,$ these equations, along with equation \reff{c22}, or equivalently \reff{c23}, complete the compatibility of constraints yielding
\eea{
\nn \frac{y_j-\a}{[\s(\a-y_j)+\b^2]} +\sum_{k=1}^L\frac{c_2 f(z_k)-c_1}{f(z_k)-y_j} &=
2\sum_{l\neq j}^M \frac{c_1 -c_2y_l}{y_j -y_l},\\
\label{c1c2eq} c_1\s &= c_2(\s\a+\b^2).
}
For the case $\b=0$,  the constraint \reff{c1c2eq} is no longer necessary. In this instance the bosonic degree of freedom decouples and we may project onto a BCS system which will be discussed in the next subsection. 

An equivalent expression for the Bethe ansatz equations is
obtained by manipulating the terms,
\eean{
\begin{split}
\frac{1}{2y_j}\left(\frac{\a}{\s(c_1/c_2-y_j)}+c_2(2M-2-L)\right) -\frac{1}{2\s(c_1/c_2-y_j)}  +\frac{1}{4}\sum_{k=1}^L&\frac{2}{y_j/c_2 -f(z_k)/c_2} \\
+\frac{c_1}{2y_j}\left(\sum_{k=1}^L\frac{1}{f(z_k)-y_j} +\sum_{l\neq j}^M \frac{2}{y_j -y_l}\right) &= \sum_{l\neq j}^M \frac{c_2}{y_j -y_l}
\end{split}\\
c_2(\s\a+\b^2)\b &=  c_1 \s\b.
}
In the special case of $\s=0=\k$ we find $c_2=0$, $\b^2 c_1 =-\b^2\conj{g(z_k)}g(z_k)=-G^2$ and the Bethe ansatz equations reduce to
\eean{
\frac{\a}{2G} -\frac{y_j/G}{2} -\frac{1}{4}\sum_{k=1}^L\frac{2}{f(z_k)/G-y_j/G} &= \frac{1}{2}\sum_{l\neq j}^M \frac{2}{y_j/G -y_l/G}.
}
This result is in complete agreement with that of the first case, where we derived the Bethe ansatz equations for $\k\neq0$ and $\s=0$ and then set $\k=0$ at the end to obtain the limiting case.


\subsection{Recovering Known Exactly Solvable Subcases}   
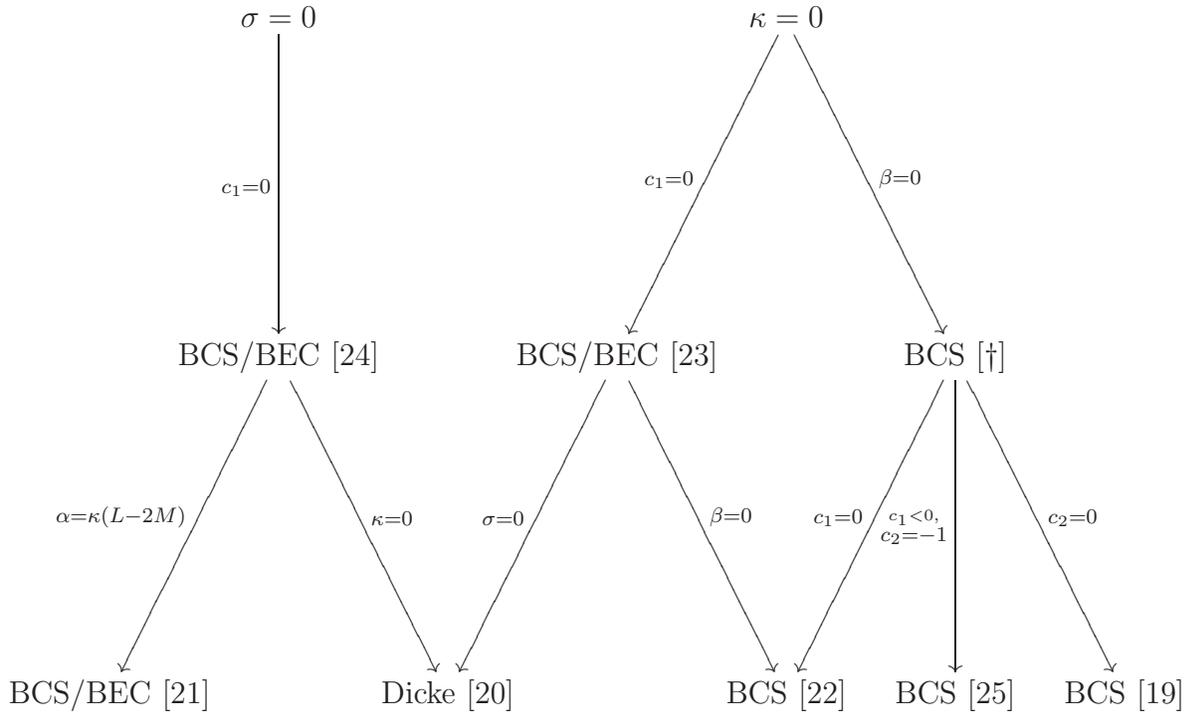
\begin{figure}[htbp]
\centering
			$ \xymatrix @!=40pt {&   \s=0  \ar[dd]_{c_1=0} &  &   &  \k=0 \ar[ddl]_{c_1=0}   \ar[ddr]^{\b=0} &  & \\
			& & & & & & \\
&  \textrm{BCS/BEC \cite{lrdo11}} \ar[ddl]_{\a=\k(L-2M)} \ar[ddr]^{\k=0} &  &  \textrm{BCS/BEC \cite{dilz11}} \ar[ddl]_{\s=0}  \ar[ddr]^{\b=0} &  &  \textrm{BCS $[\dagger]$} \ar[ddl]_{c_1=0} \ar[dd]_{\stackrel{c_1<0,}{c_2=-1}}  \ar[ddr]^{c_2=0} &\\
		&	& & & & &\\
 \textrm{BCS/BEC \cite{dhl06}} & &  \textrm{Dicke \cite{g76}}  & &  \textrm{BCS \cite{ilsz09}} & \textrm{BCS \cite{dlrrr11}}  &  \textrm{BCS \cite{r63}} 
} $
\caption{Hierarchy of known exactly solvable models that are limiting cases of our
solutions for $\sigma=0$ and $\kappa=0$. Key: \cite{lrdo11} is the 2-channel $p+ip$-wave
BCS coupled to bosonic molecular pairs with self-interaction term and no pair-pair interaction; 
\cite{dilz11} is the $p+ip$-wave BCS coupled to bosonic molecular
pairs with no self-interaction term;  \cite{dhl06} is a BCS system coupled to a bosonic mode which
was derived by use of Sklyanin's boundary QISM; \cite{g76} is equivalent to the Dicke
model as studied by Gaudin; \cite{ilsz09} is the $p+ip$-wave BCS model; \cite{dlrrr11} is
a BCS model with an energy cut-off used for the study of heavy nuclei; \cite{r63} is the
original reduced $s$-wave BCS case due to Richardson; $[\dagger]$ is a general exactly
solvable BCS model which it appears has not previously been explicitly identified in the
literature.}
	\label{fig:submodels}
\end{figure}

In the above subsections we determined manifolds in the coupling parameters of \reff{genHAM} for which an exact solution exists\footnote{While we have tried to be general in deriving this manifold, it is potentially possible to relax some of the assumptions made on the functions introduced in Section \ref{sec:constraints}. This will be considered further in future work.}. 
%
%
Taking appropriate limits of the general exactly solvable models yields eight subcases which we have presented in Figure \ref{fig:submodels}. Seven of these subcases are known \cite{r63,lrdo11,dilz11,dhl06,g76,ilsz09,dlrrr11}\footnote{In some of the subcases a change of variable is required to bring the Hamiltonian into the form cited.}. 


The three tiers apparent in the graph in Figure \ref{fig:submodels} correspond to the number of free parameters in the solvable models which determine the coupling interaction strengths and the functional relationship between $f(z)$ and $g(z)$. The top tier consists of the most general exactly solvable models which were derived in sections \ref{sec:case1} and \ref{sec:case2}. These models each have 3 free parameters up to an arbitrary energy rescaling. Models in the middle tier have 2 free parameters and models in lowest tier are described by only 1 free parameter.
%

\vfill
\newpage

For the $\s=0$ case the Hamiltonian is of the form
\eean{ H = \a N_0 +\k N_0^2 +\sum_{k=1}^L f(z_k)N_k -\b \sum_{k=1}^L \left(g(z_k)b_0 b_k\dag +\conj{g(z_k)} b_0\dag b_k\right),
}
and is exactly solvable with eigenstates
\eean{
\ket{\Psi} = \prod_{j=1}^M \g(y_j)\left(b_0\dag + \sum_{k=1}^L \frac{\b g(z_k)}{f(z_k)-y_j}b_k\dag\right)\ket{0}
}
when 
\eea{
\label{BCSsym1} \b^2 |g(z_k)|^2 +c_1&= c_2 f(z_k),~~~c_2=\k\\
\nn y_j-(\a+\k) +\sum_{k=1}^L\frac{ \k f(z_k)- c_1}{f(z_k)-y_j} &= 2\sum_{l\neq j}^M \frac{c_1-\k y_l}{y_j-y_l}.
}
Equation \reff{BCSsym1} affects the pairing symmetry in the BCS part of the model. The 3 independent parameters in this model are $\b$, $\k$, and $c_1$.

For the $\k=0$ case the Hamiltonian is of the form
\eean{ H = \a N_0 + \sum_{k=1}^L f(z_k)N_k -\b \sum_{k=1}^L \left(g(z_k)b_0 b_k\dag +\conj{g(z_k)} b_0\dag b_k\right) -\s\sum_{k,s}^L g(z_k)\conj{g(z_s)}b_k\dag b_s,
}
and is exactly solvable with eigenstates
\eean{
\ket{\Psi} = \prod_{j=1}^M \g(y_j)\left(b_0\dag + \sum_{k=1}^L \frac{(c_2-c_1y_j)g(z_k)}{c_1\b(f(z_k)-y_j)}b_k\dag\right)\ket{0}
}
when 
\eea{
\nn c_2(\s\a+\b^2)&=c_1\s,\\
\label{BCSsym2} |g(z_k)|^2 +c_1&= c_2 f(z_k),\\
\nn \frac{c_1(y_j-\a)}{c_2-c_1y_j} +\sum_{k=1}^L\frac{c_2 f(z_k)-c_1}{f(z_k)-y_j} &= 2\sum_{l\neq j}^M \frac{c_1 -c_2y_l}{y_j -y_l}.
}
Immediately we notice the similarity with the $\s=0$ case in that the equation \reff{BCSsym2} affects the pairing symmetry in the BCS part of the model. 

In the limit of $\b=0$, the $\k=0$ model reduces to a model with no coupling between the bosonic degree of freedom and the Cooper pairs. Here we can simply project out the bosonic degree fo freedom from the Hilbert space of states and consider the Hamiltonian as BCS type only.
The BCS model in this case has neither $p+ip$-wave or $s$-wave symmetry, but a generalisation of both. The Hamiltonian is of the form
\eea{
\label{newBCSham} H_{BCS} &= \sum_{k=1}^L f(z_k)N_k -\s\sum_{k,s}^L g(z_k)\conj{g(z_s)}b_k\dag b_s,
}
and is exactly solvable with eigenstates\footnote{
These eigenstates are obtained by renormalising the eigenstate \reff{ansatz} to accomodate for the factor $\prod_{j=1}^M\g(y_j)$ which approaches zero in the limit $\b\to0$.}
\eean{
\ket{\Psi} = \prod_{j=1}^M \left(\sum_{k=1}^L \frac{(c_2-c_1y_j)g(z_k)}{f(z_k)-y_j}b_k\dag\right)\ket{0}
}
where 
\eea{
\label{newBCShamBA1} |g(z_k)|^2 +c_1&= c_2 f(z_k),\\
\label{newBCShamBA2} -\frac{1}{\s} +\sum_{k=1}^L\frac{c_2 f(z_k)-c_1}{f(z_k)-y_j} &= 2\sum_{l\neq j}^M \frac{c_1 -c_2y_l}{y_j -y_l}.
}
Here $c_1=0$ results in $p+ip$-wave symmetry, while $c_2=0$ results in $s$-wave symmetry\footnote{Strictly speaking we also have to impose that the pairing is triplet type in the $p+ip$-wave case and singlet type in the $s$-wave case.}.

We take this moment to reflect on some of the historical passage of events in integrable pairing Hamiltonians. Although the construction of exact eigenstates in the $s$-wave model traces back to Richardson's work of 1963, it was only in 1997 that the conserved operators were constructed \cite{crs}. It was soon realised that these operators could be reproduced following the approach of Gaudin \cite{g76} by considering the rational class of Gaudin magnets. By extending this to the trigonometric and hyperbolic classes more general integrable Hamiltonians were constructed in 2001 \cite{ado01,des01}. It turns out that the conserved operators in the hyperbolic case are those of the $p+ip$-wave Hamiltonian, a model which came to prominence through the work of Read and Green in 2000 \cite{rg00}. However it was not immediately apparent that the hyperbolic class of conserved operators could be combined to produce the $p+ip$-wave Hamiltonian, and it was only in 2010 that the explicit relationship was made \cite{dilsz10}.

It is somewhat surprising that the above results show that the rational and hyperbolic cases can be seen as two limits of a more general pairing Hamiltonian (\ref{newBCSham}) with the constraint (\ref{newBCShamBA1}) and the Bethe ansatz equations (\ref{newBCShamBA2}). 
The number of free parameters is the same in both limiting Hamiltonians. This is in some contrast to the case of the $XXX$ spin chain, which is the rational limit of the $XXZ$ spin chain obtained by a one-variable reduction in the coupling co-efficient of the $S_i^zS_{i+1}^z$ interactions.   

\section{Conclusion}

We introduced a variational approach for the Quantum Inverse Scattering Method to exactly solve a class of Hamiltonians via Bethe ansatz methods. The procedure was conducted in the framework of variational Hamiltonians describing BCS-BEC crossover physics through interacting Cooper pairs and a bosonic degree of freedom. We obtained general exact solvability requirements which included seven subcases which have previously appeared in the literature.  An initial question to consider is whether the general forms of exactly solvable Hamiltionians, in Subsection 2.1.1 and 2.1.2 respectively, do admit a  set of conserved operators. We have found that such a set is indeed constructible, either through the Gaudin algebra approach of \cite{osdr05} (by relaxing constraints imposed in Subsection 3.1 of their work), or by a suitable adaptation of the classical Yang-Baxter equation approach of \cite{s09}. However we emphasize that our motivation was to undertake calculations to obtain exactly solvable Hamiltonians in a manner which did not rely on any prior knowledge of integrability through the existence of a set of conserved operators.     

Finally, we remark that there remains scope to further extend this approach to a more general level than that which we have considered here. For example the exactly solvable Russian doll BCS model \cite{lrs04,dl04} does not fit into the above scheme, both in that the pair-pair interaction is not factorisable, and the wave-function ansatz is of a different type. Furthermore pairing Hamiltonians such as those in \cite{ado01,des01,s09} contain interaction terms which are not present in our starting Hamiltonian (\ref{genHAM}), and similar extended exactly solvable models can also be constructed within the above approach.

\subsubsection*{Acknowledgements} This work was supported in part by the Australian Research Council under Discovery Project DP110101414. A. Birrell acknowledges the support of an Australian Postgraduate Award.

\end{document}